\documentclass[letter]{ptptex}

\usepackage{graphicx}
\usepackage{wrapft}

\newcommand{\vect}[1]{\mbox{\boldmath${#1}$}}
\newcommand{\vex}{{\vect x}}
\newcommand{\beqa}{\begin{eqnarray}} \newcommand{\eeqa}{\end{eqnarray}}
\newcommand{\bseqa}{\begin{subeqnarray}} \newcommand{\eseqa}{\end{subeqnarray}}
\newcommand{\beq}{\begin{equation}} \newcommand{\eeq}{\end{equation}}

\notypesetlogo                       %comment in if to eliminate PTPTeX 
\title{%    %You can use \\ for explicit line-break
A Calculation of the Viscosity to Entropy Ratio of a Hadronic Gas     %
}

\author{%       %Use \scshape  for the family name
Shin \textsc{Muroya}$^{1)}$ and Nobuo \textsc{Sasaki}$^{2)}$%
}

\inst{%         %Affiliation, neglected when [addenda] or [errata]
$^1$ Tokuyama University, Shunan 745-8566 Japan \\
$^2$ Research Center for Nanodevices and Systems, \\
Hiroshima University, Higashi-Hiroshima 739-8527 Japan 
}

\recdate{August 24, 2004}%            %Editorial Office will fill in this.

\abst{%         %this abstract is neglected when [addenda] or [errata]
We calculate thermodynamic quantities and transport coefficients of a
hadronic gas using a Monte Carlo hadronic 
collision event generator URASiMA.  
The obtained shear viscosity to entropy density ratio of the hadronic gas is 
approximately 0.3 -- 0.6 in the range $T$ = 100 -- 180MeV and is quite
insensitive to the baryon number density. 
}

\begin{document}

\maketitle

\section{Introduction}

Hydrodynamical models are among the most thoroughly established models of 
multiple production phenomena\cite{F-L}. In particular for RHIC experiments, 
hydrodynamical models are accepted as the most successful models 
to reproduce $v_2$ data\cite{QM04}.  To this time, in 
almost all hydrodynamical models,
viscosities and heat conductivity have been ignored for simplicity.
Teaney\cite{Teaney} attempted to estimate the size of the effect 
that may appear 
in the present 
hydrodynamical models applied to  RHIC experiment in the case that the 
shear viscosity is 
taken into account, and he concluded that 
the viscous term in the Navier-Stokes equation must be smaller 
than unity in order to retain good agreement between the blast 
wave model with viscous corrections and experimental results.  
A peculiarly small ratio of the viscosity 
to entropy has been identified as a possible special feature of the QGP liquid
\cite{Nakamura}.
Here we evaluate the shear viscosity and entropy for a hadronic gas.

Calculating macroscopic material constants on the basis of 
a microscopic picture is one of 
 the most important tasks in theoretical physics. 
This is, however, very difficult in general.  Particularly in the case of 
strongly interacting hadrons, there exists no perturbation theory 
 and  no systematic method to calculate the relevant quantities. 

In this letter, we evaluate the transport coefficients of dense, 
hot hadronic matter employing  a hadro-molecular dynamic 
calculation that an event generator URASiMA
(Ultra-Relativistic AA collision Simulator based on Multiple
Scattering Algorithm)\cite{Kumagai}. 
The obtained heat conductivity and shear viscosity 
exhibit a  temperature dependence of $\sim T^{5}$, which is 
stronger than the result obtained from naive dimensional 
analysis, $\sim T^{3}$. Muronga 
attempted a similar calculation for a pure meson gas using UrQMD, 
in which case, several interactions must be switched off by hand in 
order to realize the stationary state\cite{Muronga}. In
the present letter, we investigate a finite baryon number system 
with all hadronic interactions, including those of baryons.

\section{Hadro-molecular dynamics}

In order to construct a statistical ensemble state, we carry out 
a molecular dynamical calculation of hadrons in a box.
The time evolution of the system is simulated with the Monte-Carlo 
collision event generator URASiMA. One of the present authors
(N.\ S.) improved and tuned the parameters of URASiMA so as to
realize an equilibrium state of hadrons for temperatures 
in the rande 100 -- 180 MeV for nuclear densities near 
twice the normal nuclear density\cite{Sasaki-ptp}. In Ref.~7), 
equilibration of the system is discussed in detail 
and thermodynamical quantities 
are analyzed.  Several kinds of diffusion constants have already
investigated on the basis of the ensembles\cite{PRC, EPL}.

The present version of the URASiMA contains the particles 
listed in the Table I.
This version is the so-called ^^ ^^ two-flavor" and ^^ ^^ low-energy" 
version, which ignores anti-baryons and strangeness. As a hadronic gas model,
this model 
covers the region characterized by temperature lower than Kaon's mass and 
chemical potential lower than the baryon mass.  The region on which we 
focus our consideration is below $T_{c}$ ($\approx$ 200 MeV) and up to 
twice the normal nuclear baryon number density. We believe that this region
is well within the regime in which the present version of URASiMA is valid.

% ----- Table 1 -------------------------------------------------------
\begin{table}[tb]
\begin{center}
  \caption{
    Baryons, mesons and their resonances included in the URASiMA.
  }
  \label{tab:plst}
  \begin{tabular}{c|cccccccc}
\hline \hline
    nucleon  & $N_{938} $ & $N_{1440}$ & $N_{1520}$ & $N_{1535}$ &
               $N_{1650}$ & $N_{1675}$ & $N_{1680}$ & $N_{1720}$ \\
               \hline
    $\Delta$ & $\Delta_{1232}$ & $\Delta_{1600}$ &
               $\Delta_{1620}$ & $\Delta_{1700}$ &
               $\Delta_{1905}$ & $\Delta_{1910}$ &
               $\Delta_{1950}$ & \\ \hline
    meson & $\pi$ & $\eta_{s}$ & $\sigma_{800}$ & $\rho_{770}$ &&&& \\
\hline
  \end{tabular}
\end{center}
\end{table}

%%%%%%%%%%%%%%%%%%%%%%%%%%%%%%%%%%%%%%%%%%%%%%%%%%%%
%  Figs.1 and 2 
\begin{figure}[b]

\begin{center}
\begin{minipage}{ 0.45\linewidth}
\includegraphics[width= 0.93\linewidth]{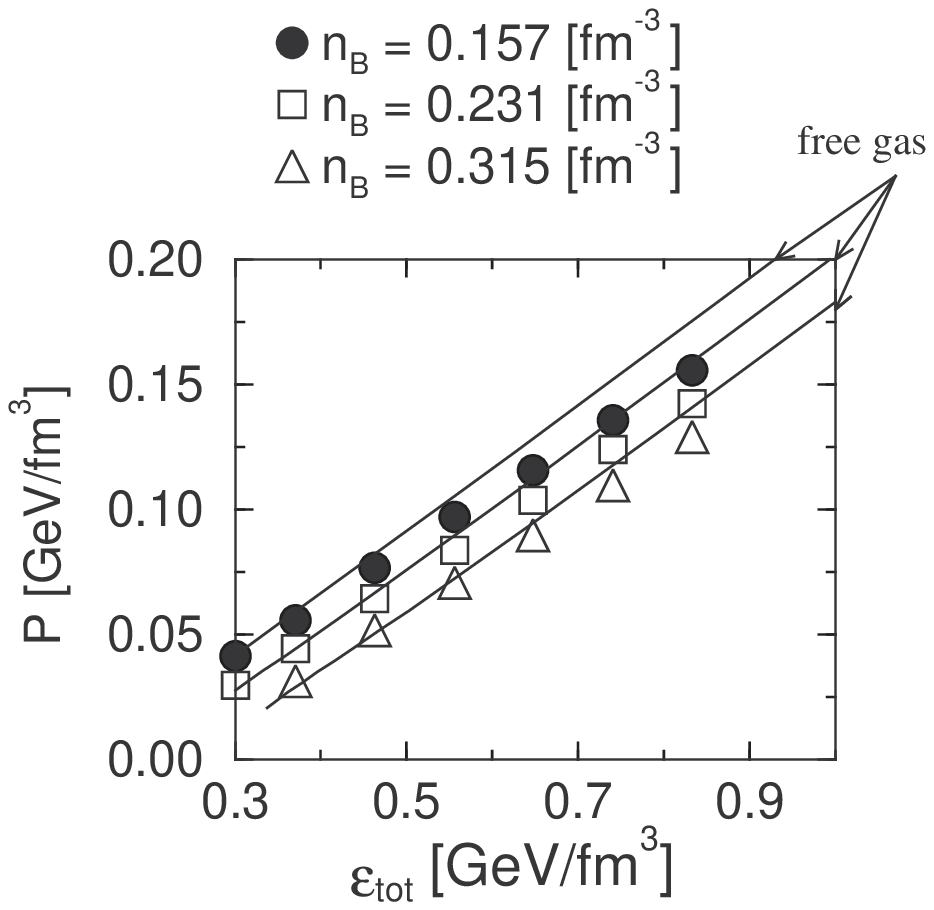}
\caption{Equation of state of a hadronic gas obtained using
URASiMA. 
}
\end{minipage}
\begin{minipage}{ 0.09\linewidth}

\end{minipage}
\begin{minipage}{ 0.45\linewidth}
\includegraphics[width= 0.95\linewidth]{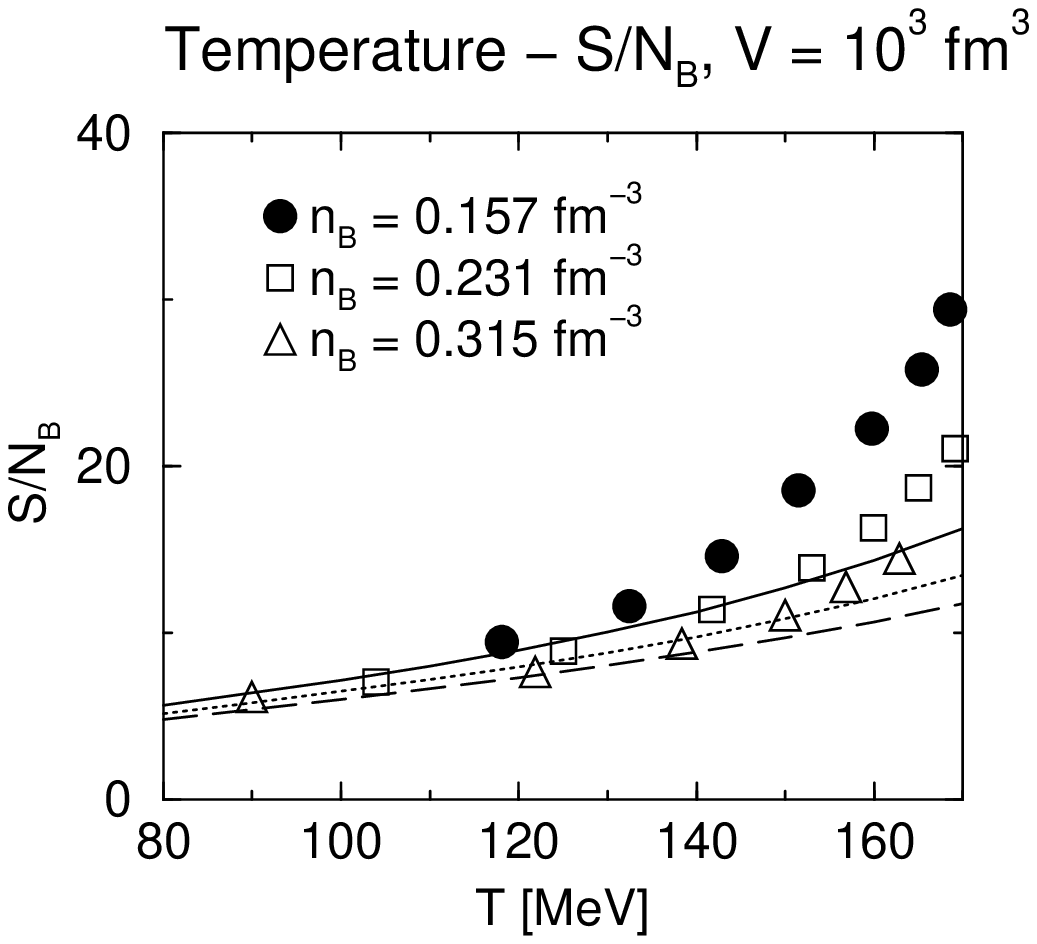}
\caption{Entropy of a hadronic gas obtained using URASiMA.
}
\end{minipage}
\end{center}
\vspace{5mm}
\label{Fig-PolSus}
\end{figure}
%%%%%%%%%%%%%%%%%%%%%%%%%%%%%%%%%%%%%%%%%%%%%%%%%%%%

Thermodynamical quantities of the systems of interest 
are  investigated in Ref.~7).
Figure 1 displays the obtained equation of state 
(the pressure $P$ as a function 
of the energy density $\varepsilon$). Throughout our simulation, 
the box size is fixed as $10^3$ fm$^3$. The black circles represent the 
normal nuclear density $n_{b0}$ ($n_{b0}=$ 0.157 (1/fm$^3$)), the 
white boxes represent
1.5 times $n_{b0}$, and the triangles represent 
2 times $n_{b0}$.
We note from Fig.\ 1 that the bulk viscosity of the
hadronic gas almost vanishes.  Because the current $P'$ which corresponds to 
the bulk viscosity is given by\cite{Zubarev}
\beq
P'=P-\frac{\partial P}{\partial \varepsilon} \varepsilon.
\eeq
$P'$ vanishes if the pressure is proportional to energy density.

Figure 2 displays the entropy density calculated in Ref.\ 7)
 as $S = \sum \rho \ln \rho $, where $\rho$ is 
{\it the ensemble averaged } 
density function and $\sum$ represents the sum 
over phase space, respectively.  
Throughout our simulation, the baryon number is conserved. 
The entropy density grows more rapidly than that  
in the free hadron model (the curves in the figure) 
for the temperatures approximately equal to or larger 
then the pion mass. 

\section{Relaxation of the currents}

According to linear response theory, a transport coefficient 
is obtained from the correlation of the corresponding current.\cite{Kubo}  
The shear viscosity and heat conductivity are obtained from the 
correlation of the 
heat current $P_{i}$ and stress tensor $\pi_{ij}$ as\cite{Nakajima}
\beqa
%\begin{array}{lcl}
\kappa &=& \lim_{\varepsilon \to +0} \frac{1}{T}\int d^3 \vex '{\int^{t}}_{-\infty} dt' { e}^{-\varepsilon(t-t')}
\langle P_{x}(t,\vex),P_{x}(t',\vex ')\rangle ,\\[0.2cm]
\eta_{s} &=& \lim_{\varepsilon \to +0} \frac{2}{T} \int d^3 \vex '{\int^{t}}_{-\infty} dt' { e}^{-\varepsilon(t-t')}
\langle \pi_{xy}(t,\vex),\pi_{xy}(t',\vex ')\rangle, \label{cor}
%\end{array}
\eeqa
where $P_{i}$ and $\pi_{ij}$ are the spatial-temporal and spatial-spatial
components of the energy-momentum tensor, respectively.  

In the hadro-molecular dynamics described by URASiMA, 
the energy-momentum tensor 
is given by the sum of that of each particle:
\beqa
T^{\mu\nu}(\vex,t)&=& \sum _{l} T^{\mu\nu}_{(l)} \delta(\vex - \vex_{(l)}(t))\\
&=& \sum _{l} \frac{p^{\mu}_{(l)}p^{\nu}_{(l)}}{p^{0}_{(l)}}\delta(\vex- \vex_{(l)}(t)).
\eeqa
Here, the subscript $l$ indexes the particles. 
Average is taken as an ensemble average:
\beq
\langle \cdots \rangle = \frac{1}{\mbox{number of ensembles}}\sum_{\mbox{ensemble}}\delta_{(l,l')}.
\label{eqn;avg}
\eeq
Correlation is assumed to exist only between the same particle in a state.\footnote{A detailed explanation of the procedure will be given in Ref.~16).}

%%%%%%%%%%%%%%%%%%%%%%%%%%%%%%%%%%%%%%%%%%%%%%%%%%%%%%%%%%%%%%%%%%%%%%%%%%
%  Fig.3  Pi-piCorrelation 
\begin{wrapfigure}{1}{7cm}
%\figurebox{60mm}{3cm}
\centerline{
\includegraphics[width=1.0 \linewidth]{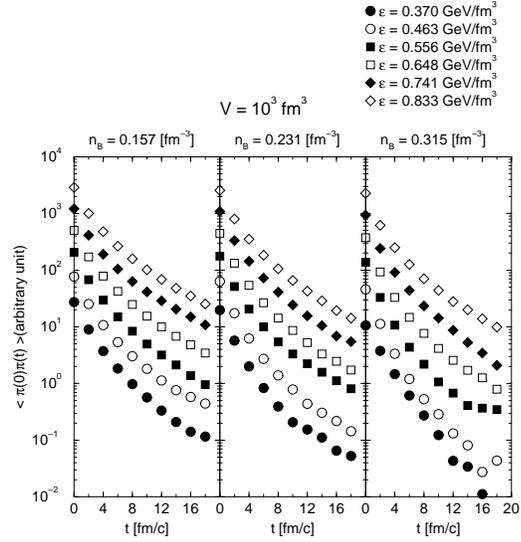}
}
\caption{Correlation function of the viscous shear tensor $\pi_{x,y}$.}
\end{wrapfigure}
%%%%%%%%%%%%%%%%%%%%%%%%%%%%%%%%%%%%%%%%%%%%%%%%%%%%%%%%%%%%%%%%%%%%%%%%%
Figures 3 displays the correlation function of the stress tensor 
$\pi_{ij}$, which represents the relaxation of the viscous 
shear tensor and appears in the integrand of Eq.~(\ref{cor}).  
At temperatures near the pion mass, mesonic degrees of freedom 
gradually come to dominate transport phenomena. Detailed analyses 
of the separate contributions of the baryonic sector and mesonic sector 
will be reported elsewhere.\cite{Sasaki3}

In each case, for values of $t$ greater than 
about $t=20$ fm, the correlation function 
becomes smaller than $\sim$ $10^{-3}$ and 
can no longer be distinguished from the noise. 
Because we are interested in phenomena 
on the hadronic scale, in actual 
calculations, we cut off the infinite integration in Eqs.~(2) and (3) 
at $t=20$ fm by hand. 
It is known that the final results are not sensitive to this cut off value.

\section{Transport coefficients}

%  Fig.4  Shear viscosity 
\begin{figure}[b]
%\begin{wrapfigure}{1}{17.0cm}
%\figurebox{60mm}{3cm}
\centerline{
\includegraphics[width=.9 \linewidth]{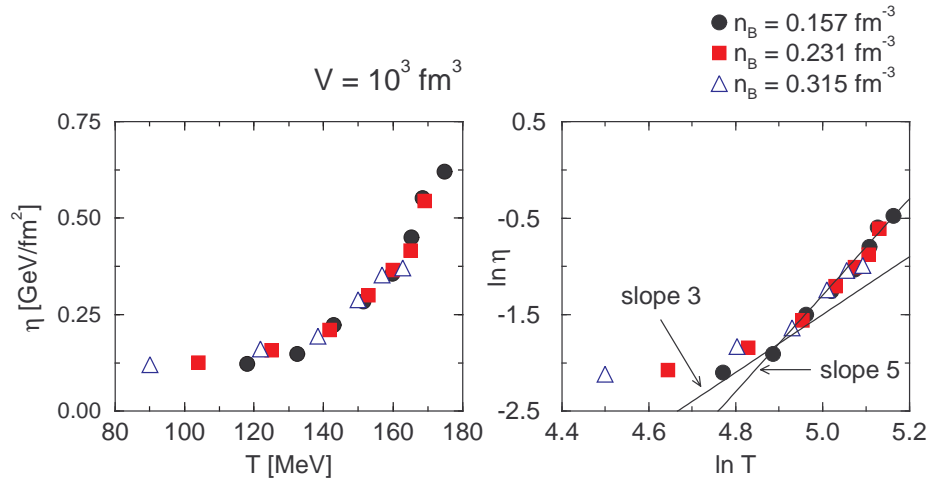}
}
\caption{Shear viscosity $\eta_{s}$ as a function  of temperature.}
%\end{wrapfigure}
\end{figure}

Figure 4 displays values of the calculated shear viscosity.
Though the temperature range is not large, 
it is seen that the temperature dependence of $\eta$ is 
$\sim T^5$, which is significantly stronger than that obtained 
from naive dimensional analysis, $\sim T^3$.
This strong temperature dependence is caused by the liberation of 
the pion degrees of freedom, because the temperature is slightly 
above the pion mass.  A similar temperature dependence also appears in
the heat conductivity shown in  Fig.~5.  The somewhat peculiar 
behavior of the heat 
conductivity in the region of very low temperature 
can be attributed to  the contribution of 
the baryons, whose mass is larger than the temperature and 
whose number must be conserved.  
A detailed investigation will be presented elsewhere.\cite{Sasaki3}

%  Fig.5  Heat conductivity 
\begin{figure}[tb]
%\begin{wrapfigure}{1}{17.0cm}
%\figurebox{60mm}{3cm}
\centerline{
\includegraphics[width=.9 \linewidth]{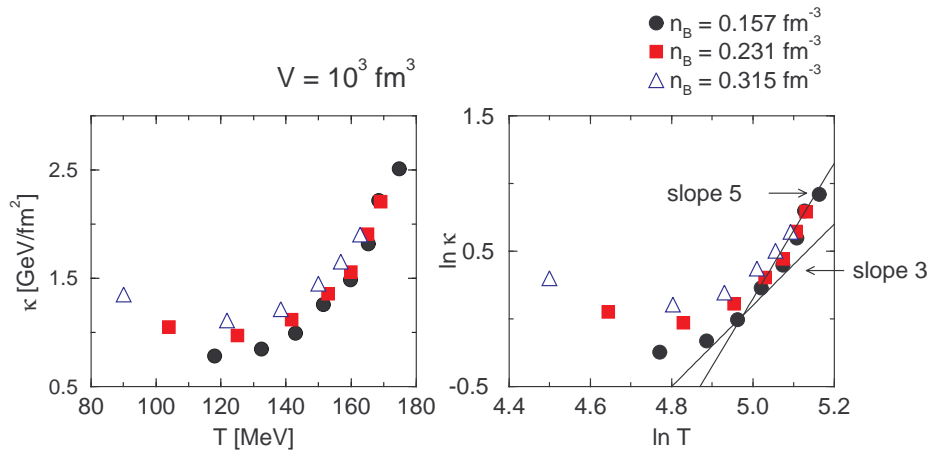}
}
\caption{Heat conductivity $\kappa$ as a function  of temperature.}
%\end{wrapfigure}
\end{figure}

The obtained shear viscosity is approximately 0.2 -- 0.6 [GeV/fm$^2$] and 
is almost independent of the baryon number density.  
Though the system treated by the URASiMA 
is not a simple pion gas, the transport of
stress is dominated by pions, because the temperature 
of the system is sufficiently smaller than the masses of hadrons other 
than pions. 

About twenty years ago, 
Gavin calculated the shear viscosity of a pion gas \cite{Gavin} 
and obtained 0.05 -- 0.15 [GeV/fm$^2$], which is 
approximately one-fourth as large as our result.
According to the recent calculation by Dobado and Llanes-Estrada, 
\cite{Dobado}
the shear viscosity of a pion gas is approximately 
0.4 [GeV$^{3}$]( $\sim$ 10 [GeV/fm$^2$]) 
if a constant scattering amplitude is assumed 
(see Fig.~2 of Ref.~13))
 and 0.004 [GeV$^{3}$]( $\sim$ 0.1 [GeV/fm$^2$]) in the simple 
analytical phase shift model (see Fig.~3 of Ref.~13)).  
Our model gives a result between these values.  For a pure meson gas, 
 using UrQMD in which several interactions are modified to equilibrate the
system, Muronga obtained a shear viscosity which is almost one-half 
of our result \cite{Muronga}.

\section{$\eta_{s}$ to entropy density ratio}

\begin{wrapfigure}{1}{6.0cm}
%\figurebox{60mm}{3cm}
\centerline{
\includegraphics[width=1.0 \linewidth]{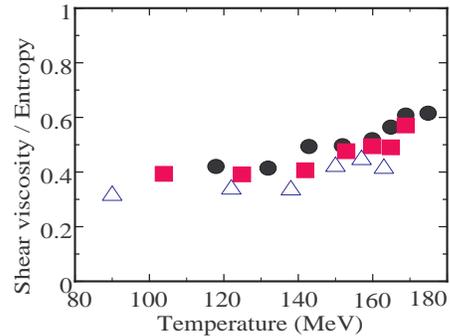}
}
\caption{Viscosity to entropy density ratio of a hadronic gas.
The black circles represent the 
normal nuclear density $n_{b0}$, the 
boxes represent
1.5 times $n_{b0}$, and the triangles represent 
2 times $n_{b0}$.}
\end{wrapfigure}

Usually, the Reynolds number $R$ is evaluated to estimate 
how large an effect is caused by viscous terms. 
If Bjorken's 1+1 dimensional Scaling 
solution is adopted as a particular solution, the inverse Reynolds 
number, $R^{-1}$, is given by\cite{Reynolds}
\beqa
R^{-1} &=& \displaystyle{\frac{\eta_{s}+\frac{2}{3}\eta_{v}}{\tau (\varepsilon +P)}}\\[0.5cm]
&=&  \displaystyle{ \frac{\eta_{s}+\frac{2}{3}\eta_{v}}{s}} \frac{1}{\tau T},
\eeqa
where $\tau$ is the proper time.

Instead of the inverse Reynolds number, which depends on the solution 
of hydrodynamics, we can use the shear viscosity to entropy density ratio 
for simplisity\cite{Nakamura}.
The shear viscosity to entropy density ratio, $\eta_{s}/s$, 
obtained in our simulation is plotted in Fig.~6.
It is seen that $\eta_{s}/s$ is approximately 0.4 
and quite insensitive to the baryon 
number density.
The peculiarly small value $\eta_{s}/s$ for QCD matter has 
been pointed out by several 
authors,\cite{Nakamura}
but even in the case of hadronic gas, the value of $\eta_{s}/s$ 
obtained in our simulation 
is approximately order unity 
or smaller. The URASiMA  consists of ordinary hadronic 
collisions only\cite{Sasaki-ptp}; that is, there exists no 
special strong correlation, 
such as strings and potentials. 
The main difference between URASiMA and ordinary non-relativistic molecular 
dynamic simulations lies in the relativistic properties of the system. 
Therefore, we believe that the result $\eta_{s}/s \le 1$ may be a common 
feature of a {\it relativistic gas}. 
More detailed analysis is now in progress.\cite{Sasaki3}

\section*{Acknowledgements}
We would like to thank Professor Teiji Kunihiro and 
Professor Atsushi Nakamura for their
continuous encouragement. Collaboration with Chiho Nonaka 
in the early stage of 
the present work was very fruitful.
Stimulating discussions at the workshop {\it Thermal
Field Theory  and Its Application} held at 
the Yukawa Institute for Theoretical Physics was 
extremely helpful. One of the authors (S.~M.) thanks  
ERI of Tokuyama University for financial support.
This work was started under the 
supervision of the late Professor Osamu Miyamura, Hiroshima University.

%\appendix
%\section{First Appendix} %Empty argument \section{} yields `Appendix'. 
%
%\section{Second Appendix}

\end{document}